\documentclass{article}
\usepackage{emulateapj}
\usepackage{float,epsfig,psfig}

\newcommand{\GA}{\mbox{\raisebox{-0.6ex}{$\stackrel{\textstyle>}{\sim}$}}}
\newcommand{\hd}{HD~163296}
\newcommand{\hh}{{Herbig-Haro}}

\newcommand{\hae}{{Herbig~Ae}}
\newcommand{\haebe}{{Herbig~Ae/Be}}
\newcommand{\cxo}{{\sl Chandra}}
\newcommand{\msun}{M$_{\odot}$}
\newcommand{\ergl}{ergs~s$^{-1}$}

\newcommand{\mdot}{$\dot{M}$}

\newcommand{\asca}{{\sl ASCA}}
\newcommand{\hst}{{\sl Hubble}}
\newcommand{\etal}{et al.}
\slugcomment{Submitted to Astrophysical Journal}

\begin{document}

\title{The Herbig Ae star HD 163296 in X-rays}

\author{
Douglas~A.~Swartz\altaffilmark{1},
Jeremy~J.~Drake\altaffilmark{2},
Ronald~F.~Elsner\altaffilmark{3},
Kajal~K.~Ghosh\altaffilmark{1},
Carol~A.~Grady\altaffilmark{4},
Edward~Wassell\altaffilmark{5},
Bruce E. Woodgate\altaffilmark{6},
Randy A. Kimble\altaffilmark{6}
}
\altaffiltext{1}{Universities Space Research Association,
NASA Marshall Space Flight Center, SD50, Huntsville, AL, USA}
\altaffiltext{2}{Smithsonian Astrophysical Observatory, MS-3, 60 Garden Street, Cambridge, MA, USA}
\altaffiltext{3}{Space Science Department,
NASA Marshall Space Flight Center, SD50, Huntsville, AL, USA}
\altaffiltext{4}{Eureka Scientific; and Laboratory for Astronomy and Solar Physics, Code 681, NASA Goddard Space Flight Center, Greenbelt, MD, USA}
\altaffiltext{5}{Thomas Aquinas College, Santa Paula, CA;
and Institute for Astrophysics \& Computational Sciences, The Catholic 
Univ. of America, Washington, DC, USA}
\altaffiltext{6}{NASA Goddard Space Flight Center, Code 681, Greenbelt, MD, USA}
\begin{abstract}

Chandra
X-ray imaging spectroscopy of the nearby Herbig Ae star HD 163296 at
100~AU angular resolution is reported.
A point-like, soft ($kT$$\sim$0.5~keV), emission-line source is detected at the location of the star with an X-ray luminosity of
$4\times 10^{29}$~\ergl\ ($\log L_{\rm X}/L_{\rm bol}=-5.48$).
In addition, faint emission
along the direction of a previously-detected Ly$\alpha$-emitting jet 
and \hh\ outflow may be present.
The relatively low luminosity, lack of a hard spectral component, and
absence of strong X-ray variability in \hd\
can be explained as originating from optically-thin shock-heated
gas accreting onto the stellar surface along magnetic field lines.
This would require a (dipole) magnetic field strength at the
surface of \hd\ of at least $\sim$100~G and perhaps as high as several~kG.
\hd\ joins the T~Tauri star TW~Hya in being the only
examples known to date of pre-main-sequence stars whose 
quiescent X-ray emission appears to be
completely dominated by accretion.
\end{abstract}

\keywords{stars: individual (HD 163296) --- stars: emission-line, Be --- stars: pre-main sequence --- X-rays: stars}

\section{Introduction}

The \haebe\ stars (Herbig 1960) are the more massive counterparts of T~Tauri 
 stars.
They are characterized by
 a strong IR excess, emission lines, and luminosity class III to V
 (Waters \& Waelkens 1998).
\haebe\ stars represent an important link between high mass stars that 
 evolve directly from embedded protostars to the zero-age main sequence
 (the onset of core H-burning) and low mass stars that stop accreting before
 reaching the main sequence and that can be observed in this
 hydrostatic pre-main sequence phase (Appenzeller 1994).

\hae\ stars are often strong X-ray emitters with luminosities of order 
 $10^{29}$ to $10^{32}$~\ergl\ (e.g., Hamaguchi \etal\ 2002; 2004). 
This high level of activity is possibly linked to the presence of an accretion  
 disk and is probably ultimately driven by magnetic fields.
However, the origin of magnetic fields in \hae\ stars is unknown. 
Main sequence A~stars lack the thick surface
 convection zones (Gilliland 1986) of late-type stars (and their T~Tauri 
 progenitors) that drive magnetic dynamo activity and power
 X-ray-emitting chromospheres and hot coronae.
Hence, they are weak X-ray emitters (Simon, Drake, \& Kim 1995; 
 Linsky 2003) or even X-ray dark (Schmitt 1997). 
Perhaps field generation in \hae\ stars is not through the solar-type $\alpha$--$\Omega$ dynamo mechanism but from 
compression of interstellar cloud fields
 in the initial collapse 
 (Dudorov \etal\ 1989), induced by deuterium-shell burning during contraction
 towards the main sequence (Palla \& Stahler 1993), or generated via
 internal differential rotation during an early accretion phase 
 (Tout \& Pringle 1995).
Such fields may decay rapidly through turbulent magnetic diffusivity as the
 radiative core develops and only be present in the youngest \hae\ stars.
 
In analogy to the T~Tauri stars, whose X-ray activity is also often much
 greater than in evolved stars of similar mass, perhaps circumstellar material
 allows other magnetic configurations such as star-disk and disk-disk fields
 to generate the strong flares, hard X-ray emission, and high X-ray luminosity
 seen in some intermediate-mass \hae\ systems.
Energetic reconnection flares may arise at the corotation 
 interface between dipole stellar fields and the inner disk 
 (Hayashi, Shibata, \& Matsumoto 1996; Shu \etal\ 1994; Birk \etal\ 2000)
 or through differential rotation of the field-threaded disks themselves
 (e.g., Romanova \etal\ 1998). 
Magnetic fields are believed to regulate protostellar collapse through 
 ambipolar diffusion, to funnel accretion from the disk, and to transfer
 disk orbital motion to collimated outflows 
 (see Feigelson \& Montmerle 1999 for a review).

\hd\ is a nearby (122 pc) \hae\ star (A1Ve, $A_V$$=$$0.^m 25$)
 with an effective temperature $T_{eff}$$=$$9300$~K, luminosity
 $\log(L_{\star}/L_{\odot})$$=$$1.48^{+0.12}_{-0.10}$, mass
 $M_{\star}=2.3 M_{\odot}$, and radius $R_{\star}=2.1 R_{\odot}$
 (van~den~Ancker, de~Winter, \& Tijn~A~Djie 1998).
Based on its position in the HR diagram,
\hd\ is at an age
 $t$$=$$4^{+6}_{-2.5}$~Myr (van~den~Ancker \etal\ 1998); 
 intermediate between heavily-embedded
 pre-main-sequence A~stars and near-ZAMS stars with debris disks.
\hd\ has a strong infrared excess (Hillenbrand \etal\ 1992; Meeus \etal\ 2001)
 and variable Balmer series emission lines
 (Baade \& Stahl 1989; Pogodin 1994; Beskrovnaya \etal\ 1998) typical of the
 Herbig Ae class though it has no associated molecular or dark
 cloud (e.g., Th\'{e} \etal\ 1985).
The large infrared excess arises from heated, optically thick dust
 within a circumstellar disk (Hillenbrand \etal\ 1992; Meeus \etal\ 2001).
The disk has a radius of $\sim$450~AU (3.\arcsec7) viewed at $\sim$60\arcdeg\
 inclination (Mannings \& Sargent 1997; Grady \etal\ 2000).
The time-variable double-peaked and P~Cygni Balmer and UV emission line 
 profiles arise from a stellar wind, an extended chromosphere, and/or 
 rotation (e.g., Catala \etal\ 1989). 
Radio continuum observations (Brown, Perez, \& Yusef-Zadeh 1993) 
 of \hd\ also suggest wind-driven mass loss.
Orthogonal to the disk is an axially-aligned chain of \hh\ nebulae extending
 several seconds of arc above and below the disk (Grady \etal\ 2000).
A Ly$\alpha$-bright jet extending $6\arcsec$
 along the SW arm of the \hh\ flow has also been discovered
 (Devine \etal\ 2000). 

\hd\ was examined with \cxo\ in order to study the spatial distribution of
 the X-radiation from the star and circumstellar environment
 (\S~\ref{s:image}) and to
 characterize the X-ray spectrum and light curve (\S~\ref{s:spectrum}).
The combination of an accretion disk and jet in a relatively old \hae\ object,
 a low extinction due to a lack of association with dark clouds,
 the presence of \hh\ objects along the jet axis,
 and an otherwise source-free field make \hd\ a
 particularly promising object for high-resolution X-ray imaging spectroscopy.
It is found that the bulk of the X-ray emission is associated with the central 
 star and its immediate surroundings.
This emission can be described as a cool, $\sim$0.5~keV, emission-line
 plasma with a luminosity $\sim$$4 \times 10^{29}$~\ergl\ in the 0.3--3.0~keV
 range.
There is also evidence for X-ray emission along the jet.
Possible origins of the X-ray emission are discussed in \S~4.

\section{X-ray Image} \label{s:image}

\hd\ was observed for 19.2~ks on 10~August 2003.
The observation utilized \cxo 's Advanced CCD Imaging Spectrometer (ACIS)
 operating in $1/4$-subarray, timed-exposure, imaging mode.
Only the central three CCDs, S2--S4, were active during the observation
 resulting in an approximately 2.\arcmin 1$\times$25.\arcmin 3 image
 and a 0.941~s frame time.
This configuration was chosen to provide
 the highest integrity image by minimizing event pileup.
Randomization of event locations at the sub-pixel level, part of the
 standard data processing pipeline, was removed by reprocessing the data.
Standard \asca\ grade 02346 events in the 0.3--8.0 keV energy range
 were included in the analysis.

\begin{figure*}[t]
\begin{center}
\includegraphics[angle=-90,width=0.60\columnwidth]{f1a.eps}
\hfil 
\includegraphics[angle=-90,width=0.60\columnwidth]{f1b.eps}
\hfil 
\includegraphics[angle=-180,width=0.60\columnwidth]{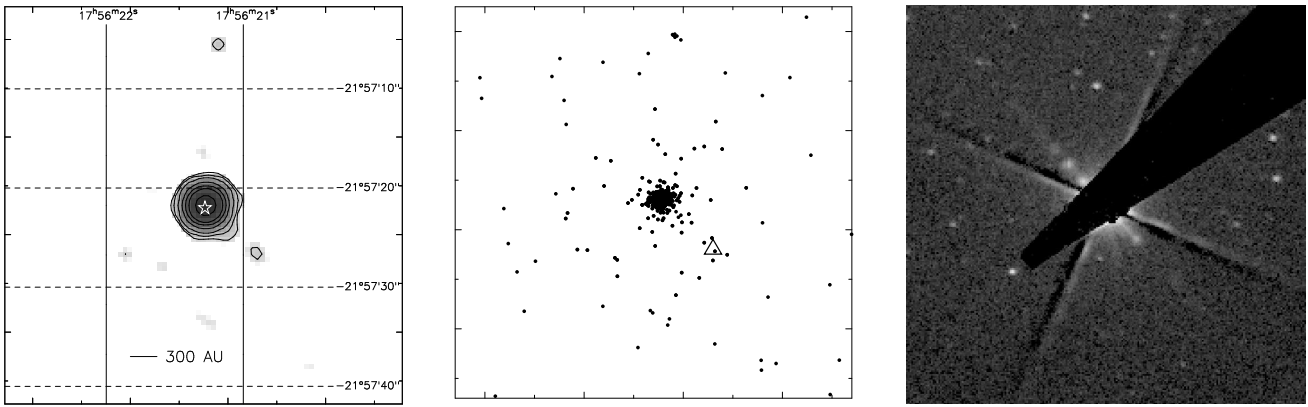}
\vspace{10pt}
\figcaption{{\em Left panel:} Smoothed 40\arcsec$\times$40\arcsec\ 
ACIS image of the 
region around \hd\ in the 0.3 to 3.0~keV bandpass obtained in 2003 August.
Tick marks denote steps of five ACIS
pixels (0.\arcsec492 pixel$^{-1}$). North is up and East is to 
the left. The star symbol marks the {\sl Hipparcos} position of \hd.
Countours denote 0.1, 0.2, 0.5, 2, 5, 10, and 30 c~pixel$^{-1}$ 
levels. The lowest contour extends $>$2\arcsec\ from the center of the
bright source along
the SW in the direction of the Ly$\alpha$ jet reported by Devine \etal\ (2000) 
(P.A. $42.\arcdeg5$$\pm$$3.\arcdeg5$).
The diffuse isolated feature is $\sim$7\arcsec\ from the central source 
 and nearly along this axis.
The debris disk (Mannings \& Sargent 1997; Grady \etal\ 2000) is oriented
perpendicular to the jet with radius $\sim$450~AU
($3.\arcsec7$) and inclination $\sim$ $60\arcdeg$.
{\em Center panel:} Individual X-ray event locations for the data shown at left.
The symbol denotes the centroid of the weak diffuse feature.
{\em Right panel:} 20\arcsec$\times$20\arcsec\ continuum-subtracted 
[S~II] image 
obtained 2004 May.
Note the occulting wedge shadow and diffraction spikes from \hd.
The jet and knot structures are clearly visible and extend $\sim$27\arcsec\
from the star along both the jet and counterjet. 
The structure in the counterjet (toward the NE) has no apparent X-ray counterparts.
The optical knots have proper motions of 0.410$\pm$0.005 sec~yr$^{-1}$.
}
\end{center}
\end{figure*}
   
Nine weak point-like sources (each with 9 to 17 counts detected corresponding to
 signal-to-noise ratios of 2.5 to 3.6) and one bright source
 (1131 counts) were detected in the field.
One of the weak sources is 0.\arcsec 20 offset in Right Ascension from a source
 in the Two Micron All Sky Survey (2MASS) 
 All-Sky Data Release catalogue.
This is comparable to the estimated statistical uncertainty in the X-ray
 position of that source.
No other astrometric catalog matches were found for any of the other
 weak X-ray sources.
The bright source is offset -0.\arcsec 26 in Dec, 0.00~s in RA,
 from the {\sl Hipparcos} position of \hd\ extrapolated to the 
 time of the X-ray observation
 ($17^{\rm h}$~$56^{\rm m}$~$21.^{\rm s}28524$,
 $-21^{\rm d}$~$57^{\rm m}$~$22.^{\rm s}0248$).
This is within the 1$\sigma$ nominal \cxo\ absolute pointing 
 uncertainty\footnote{http://cxc.harvard.edu/cal/ASPECT/celmon/}
 of 0.\arcsec 38 and infers the X-ray emission is from \hd\ and not from
 a nearby late-type companion.

\hst\ observations also place tight constraints on  
 a potential companion.
Late-type companions produce 
 conspicuous diffraction patterns in STIS chronographic imagery with 
 spatial intensity profiles distinctly different from those of early
 A~stars (see Grady \etal\ 2004a for examples in the field of HD~104237).
STIS imagery of \hd\ shows no evidence for a late-type companion 
 beyond 0.\arcsec 05 of \hd.
In addition,
 long-slit STIS spectra of \hd\ are characteristic of chromospheric and 
 transition region spectra of early-type stars and display the spatial 
 profile of an unresolved point source.
The emission line features seen in these and in FUSE data (Grady \etal\ 2004b)
 are uniformly broad and often double-peaked.
Such features are characteristic of \hae\ stars and are much broader than
 those of late-type stars; even rapid rotators like the M1Ve star AU~Mic
 (Pagano \etal\ 2000; Robinson \etal\ 2001).
Both line and UV/FUV emission are spatially coincident, at the resolution
 of \hst, with the photosphere of \hd.
In particular, emission associated with transition region temperature
 plasma, N~V and O~VI, originate on the rapidly rotating \hd\ rather
 than on a more slowly rotating, but otherwise unseen, late-type companion.
We conclude there are no late-type stellar companions to \hd.

The central 40\arcsec $\times$40\arcsec\
 region of the X-ray image is shown in the left panel of Figure~1.
A nearest-neighbor (boxcar) smoothing has been applied to the image
 and contours have been added
 to emphasize the weak extension from the source toward the SW
 and diffuse structure located $\sim$7\arcsec\ from the bright source
 along this jet axis.
The roll angle for the observation is 268.\arcdeg 5 placing any readout 
 streak along an east-west line (left-right in the figures) through 
 the bright source or roughly 45\arcdeg\ from the jet axis.
Thus, there is no contribution to the X-ray extension along the jet axis from 
 the readout streak. 
The center panel of
Figure~1 shows the positions of the individual X-ray events with the location
 of the weak diffuse feature marked. 

The right panel of Figure~1 shows, for comparison, a portion of an
 ultra-narrowband [SII] image of \hd\ taken on 2004 May 17 using the 
 Goddard Fabry-Perot (GFP) instrument and a coronagraphic wedge to occult 
 the central starlight on the 3.5m telescope at the Apache Point 
 Observatory in Sunspot, New Mexico. 
The image was constructed from an on-band
 image centered at 6724 \AA\ with a 12 \AA\ FWHM (or 540 km/s, thereby 
 sampling both the jet  at -300 km~s$^{-1}$ and the counterjet at 
 $+$340 km~s$^{-1}$) from which an off-band image centered at
 6750~\AA\ with 12~\AA\ FWHM was subtracted.
Both images are 900~s exposures taken under conditions of 1.\arcsec5 seeing. 
The GFP field of view is a circular aperture of 3.\arcmin7 diameter and 
the CCD plate scale is 0.\arcsec366~pixel$^{-1}$.
A 40\arcsec $\times$40\arcsec\ subimage of
 the full GFP image centered on the star is shown.
   Herbig-Haro knots associated with the jet and counter
jet of \hd\ are visible in the raw GFP observation,
and can be traced out to $\pm$27\arcsec\ from the star after subtraction of
the off-band image. 
The knots in both the jet and counterjet were measured at S/N$=$10. 
The weak X-ray feature does not overlap any of the optical knots when their
 positions are devolved to the time of the X-ray observation using the
 known knot proper motions.
 
The spatial distribution of the X-ray events from the central portion of
 \hd\ was compared to a
 model point spread function (PSF) to test for possible source extension.
The model PSF was generated using the
 Chandra Ray Tracer (ChaRT) web-based
 interface\footnote{http://asc.harvard.edu/chart}
 to the SAOsac raytrace code using
 the best-fit single-$kT$ model spectrum ($kT \sim 0.5$ keV, \S~3).
The radial profile (averaged over azimuth) of the data shows a slight
 excess above the model PSF in an annulus
 between 0.5 and $\sim$2.\arcsec 5 of the emission peak. 
However, the ray-trace software does not account for spacecraft dither and
 the observed excess is consistent with this model limitation. 
Therefore, except for emission along the jet, 
 there is no conclusive evidence that the bright source is extended.

\section{X-ray Spectra \& Timing} \label{s:spectrum}

\subsection{\hd}

A time-dependent
 gain\footnote{http://hea-www.harvard.edu/\~alexey/acis/tgain}
 correction was applied to
 the event file.
The spectrum was then extracted from a
 2-pixel-radius region about the center of the bright source.
Corresponding redistribution matrix and auxilliary response
 files were generated using CXC/CIAO tools (v.2.3).
Models were fit to the spectrum using the XSPEC (v.11.2) spectral-fitting
 package.
The {\tt acisabs} multiplicative absorption
 model\footnote{http://www.astro.psu.edu/users/chartas/xcontdir/xcont.html}
 was included to account for the temporal decrease in low-energy
 sensitivity of ACIS.
The C-statistic fit-minimization algorithm was used to determine the
 best-fit parameters for the unbinned data.

Figure 2 shows the spectrum, model, and fit residuals
 for an absorbed variable-abundance ({\tt vmekal}) emission-line model
 (C-statistic 214.5 for 184 spectral bins).
For display purposes, up to 5 channels have been combined
 to achieve a minimum 5$\sigma$ significance.
The parameter estimates and 90\% confidence ranges for this model are:
 hydrogen column density, $N_H/10^{20}$$=$$7.62\pm1.85$~cm$^{-2}$;
 plasma temperature, $kT$$=$$0.49\pm0.03$~keV;
 and emission measure, $EM$$=$$(1.1\pm0.3)\times 10^{53}$~cm$^{-3}$.
The metal abundances were allowed to vary in the model fits but were  
restricted to preserve solar ratios for certain element groups whose
first ionization potentials (FIPs) are similar (to allow for possible
FIP-based chemical fractionation commonly observed in late-type
stellar coronae) and whose principal X-ray
lines are formed at similar energies;
 namely Fe-Ni-Ca, Ne-Ar, Mg-Si-S-Al-Na, and C-N-O. 
In this way, the number of free parameters is not excessive. 
The estimated best fit metal abundances relative to solar are 
 0.13$\pm$0.02 for the Fe group,
 0.05$\pm$0.05 for the Ne group,
 0.18$\pm$0.08 for the Mg group, and
 0.25$\pm$0.12 for the C group.

\begin{center}
\includegraphics[angle=-90,width=\columnwidth]{f2.eps}
\figcaption{{\em Upper panel:} Observed spectrum of \hd\ (symbols) and
best-fit single-temperature thermal emission-line ({\tt vmekal}) model
(solid line). {\em Lower panel:} Fit residuals.}
\end{center}

The best-fit value of $N_H$ is slightly higher than estimated from
 the optical extinction, $A_V$=0.25 mag, assuming a normal extinction law
 (e.g., Bohlin, Savage, \& Drake 1978).
The model luminosity is $2.9 \times 10^{29}$~\ergl\
 corresponding to an intrinsic (absorption-corrected) luminosity of
 $L_{\rm X}$$=$$4.0 \times 10^{29}$~\ergl\ in the 0.3--3.0~keV energy range
 after correcting for the finite spectral-extraction region.

Additional models were applied to try to improve the fits and to 
 better characterize the composition and intrinsic absorption column.
These included relaxing the metal abundance restrictions in the original model,
 2-- or 3--temperature {\tt vmekal} models, 
 and variable-abundance absorbing columns.
None of these models gave statistically significant improvements.
The temperature and 
 overall abundance are constrained by Fe which emits numerous strong
 Fe~XVII lines in the $\sim$0.7 to 1.0~keV band.
A lower-temperature plasma (in ionization equilibrium) would be needed to
 produce a stronger O~VII triplet at $\sim$0.57~keV
 and H-like N Ly$\alpha$ at 0.5~keV; merely increasing O and N abundances
 does not alter the model spectrum.
If the abundances of Mg and Si are ungrouped and increased to 
 $\sim$0.2$Z_{\odot}$,
 then the emission at 1.35~keV (He-like Mg) and 1.84~keV (He-like Si) 
 increases but the fit is not statistically improved.
Ne~X
Ly$\alpha$ also contributes at 1~keV but the best-fit Ne abundance is
slightly lower than for other elements at 0.05 times the solar
photospheric value.

We also investigated models with an additional thermal component,
though these models resulted in no significant improvement to the
model fit for the case of both lower and higher temperatures.
Forcing a component at a fixed temperature of 3 keV 
resulted in a contribution of at most 10\% to the total flux.

Accretion shocks or shocks from a stellar wind might contribute to the
 observed X-ray emission from \hd\ (\S~\ref{s:discuss}).
Models based on the physics of shock-heated plasma (e.g. {\tt pshock} and
 variants following the work of Borkowski \etal\ 2001), 
 were also fit to the spectrum.
The resulting fits were no better than the simple absorbed {\tt vmekal}
 model.
The best-fit parameters of the shocked gas models
 (absorbing column, post-shock temperature, abundance, and emission measure)
 were comparable to the values determined from the emission-line model.

\subsection{X-ray Knot }

There are 5 counts in the 0.3--3.0 keV band
 detected within a 2.\arcsec 5 radius around the
 position of the ``knot'' shown in Figure~1.
This feature was not detected with our source-finding algorithm.
However, the expected background is only 0.6 c
 giving a probablity of 0.0004 that 5 or more Poisson-distributed counts 
 would be detected in a given region of this size. 
On the entire S3 chip (in 1/4-subarray mode) there are 7 of 2625 regions 
 of this size found to contain 5 or more counts. 
Thus the distribution is not strictly Poissonian; there are also real sources 
 present below the detect limit.
There are 13.75 regions of this size within the area of the optical jet  
 structure extending $\pm$27\arcsec\ from \hd.
Within this area, the probability of detecting a 5-count feature in a 
 2.\arcsec 5 radius circle is therefore 0.04.

The source counts are clustered near 0.8 keV, corresponding to the peak
 of the spectrum of a thermal plasma of
 temperature $\sim$0.3~keV when folded with the \cxo\ response.
The corresponding unabsorbed X-ray luminosity of the knot is 
 $\sim$ $3\times 10^{27}$~\ergl.

\subsection{\hd\ Light Curve}

The 20~ks \cxo\ observation corresponds to about 1/3 of the 
 rotation period of \hd\ (van~den~Ancker \etal\ 1998).
No flaring of the source was observed over the short observation.
The binned light curve of \hd\ (Figure~3) is consistent with a constant flux 
 ($\chi^2$=31.9 for 25~dof) with variations not more than $\sim$20\% in 
 amplitude.
A Kolomogorov-Smirnov statistic was also computed to test the source for
 time variability by comparing the cumulative event arrival times
 binned at the frame time of 0.94~s to that expected for a steady source.
This test showed the source to be marginally variable 
 (significance level 0.006) because of the shallow dip in
 the light curve about mid-way through the observation.
These two tests indicate there is no evidence for strong variability 
 during the observation.

\begin{center}
\includegraphics[angle=-90,width=\columnwidth]{f3.eps}
\figcaption{X-ray light curve of \hd\ accumulated into 800~s bins.
Data was extracted from a 2-pixel radius about the peak of the spatial
distribution of events. Only events in the 0.3--3.0 keV energy range are
included. The background contributes less than one event to the 
$>$1000 events in the light curve.}
\end{center}

\section{Discussion} 

\subsection{\hd} \label{s:discuss}

The \cxo\ observation reveals a point-like object within 0.\arcsec25 (30 AU) of
 the {\em Hipparcos} position of \hd.
The X-ray luminosity is
 4$\times$$10^{29}$~\ergl\ ($\log L_{\rm X}/L_{\rm bol}=-5.48$)
 and the best-fit temperature is 0.5 keV (6$\times$$10^6$~K).
Notably, there is no evidence for highly-variable, hot ($kT$\GA 3~keV), 
 and luminous (up to $\sim$$10^{32}$~\ergl) X-ray emission 
 that has been attributed to reconnection flares in other \hae\ systems
 (e.g., Hamaguchi \etal\ 2002; 2004) in analogy to the late-type T~Tauri stars.
The spatial coincidence, X-ray spectrum, and lack of X-ray flares strongly
 suggest the X-ray emission is associated with \hd\ and not with 
 a hidden late-type companion. 
This conclusion is further supported by the lack of spatial and spectral 
 evidence from \hst/STIS imagery
 for a late-type companion to \hd.
 
Line-driven winds also cannot account for the X-ray emission. 
The wind mass-loss rate is 
\mdot $\sim$$10^{-15}(L_{\star}/L_{\odot})^2
 (M_{\star}/M_{\odot})^{-1}$~\msun~yr$^{-1}$ (see Abbott 1982).
For \hd, \mdot$\sim$$5.5\times 10^{-13}$~\msun~yr$^{-1}$ 
 and there is too little kinetic power in the wind,
 $E$$=$$(1/2)\dot{M} v^2$$\sim$$6 \times 10^{28}$~\ergl, 
 to account for the X-ray luminosity
 even if all this energy were converted to X-radiation through shocks.
 
The presence of a circumstellar debris disk, jets, and \hh\ outflows
 are all strong indications of accretion and accretion-induced 
 outflow in \hd.
The luminosity released in accretion is 
 $L_{\rm acc}$$=$$GM_{\star}\dot{M}/2R_{\star}$ 
 (excluding the portion viscously dissipated in the disk). 
For \hd, $L_{\rm acc}$$=$6.6$\times$$10^{33}$$\dot{M}_7$~\ergl\ 
 where $\dot{M}_7$ is the mass accretion rate in units of 
 $10^{-7}$~\msun~yr$^{-1}$. 

In the absence of strong magnetic fields, the accretion energy is 
 released in an optically-thick boundary layer at the surface of the star. 
In this case, the accretion energy radiates as a blackbody from an area
 $\sim$$4\pi R_{\star}H$, where $H$$\sim$0.1$R_{\star}$ is the height of 
 the emission region (Popham \etal\ 1993), with a temperature 
 $T$$\sim$$(GM_{\star}\dot{M}/8\pi R_{\star}^2 H \sigma)^{1/4}$ or
 $\sim$7900$\dot{M}_7^{1/4}$~K for \hd.
Boundary layers will therefore be copious UV emitters but cannot 
 account for the X-ray spectrum of \hd.  

Alternatively, the inner regions of the disk could be disrupted by 
 a magnetic field associated with the accreting star. 
In this case, disk material couples to the field lines and is channeled
 onto small regions near the magnetic poles at an angle nearly normal 
 to the stellar surface
 (K\"{o}nigl 1991, see, e.g., von~Rekowski \& Brandenburg 2004 for a more 
 recent study).
Accreted material is strongly shocked on impact.
Applying the jump conditions for a shock perpendicular to the flow implies a 
 post-shock temperature $T_s=3GM_{\star}\mu m_p/8k R_{\star}$ 
 where $\mu$$=$0.6 for ionized solar abundances, $m_p$ is the proton
 mass, and the velocity just upstream of the shock is approximately the 
 free-fall value, $v_{ff}$$=$$(2GM_{\star}/R_{\star})^{1/2}$, at the surface.
For \hd, $T_s$$\sim$5.7$\times$$10^{6}$~K~---~consistent with the observed 
 X-ray temperature. 

If the accretion rate is high, however, then the stream can be of
sufficiently high density that it 
will penetrate the stellar photosphere.  Much of the
X-radiation is then attenuated by preshock gas within the accretion
flow or reprocessed within the stellar photosphere (e.g., Ulrich 1976;
Stahler, Shu, \& Taam 1980; Lamzin 1995; Calvet \& Gullbring 1998;
Lamzin 1999; Drake 2004).  The radiation field will become optically
thick with an effective temperature of
$T$$\sim$$(GM_{\star}\dot{M}/4\pi R_{\star}^3 f \sigma)^{1/4}$ where
$f$ is the fraction of the surface area covered by the accretion
stream.  For $f=0.06$ (Shang \etal\ 2002), the effective temperature
for \hd\ is only $\sim$6000$\dot{M}_7^{1/4}$~K.  

The depth in the atmosphere at
which the accretion shock will form is given by the balance
between the 
ram pressure of the stream, $p_{ram}=2\dot{M}v_{ff}/4\pi R_{\star}^2 f$,
and the photospheric gas pressure.
For typical gas pressures of a few $10^2$-$10^3$~dyne~cm$^{-2}$, the
mass accretion rate must be below $\dot{M}_7$$\sim$0.1 for the
X-ray emitting shock to lie either above or only partially within
photospheric layers. 
For significantly lower
filling factors of less than 1~\%\ favoured by some analyses of
accreting T~Tauri stars (e.g.\ Calvet \& Gulbring 1998), the X-ray
shock can remain completely buried in the photosphere even for accretion rates
as low as $\dot{M}_7$$\sim$0.01.
Thus, the observed
X-rays from \hd\ can only be consistent with an accretion scenario if
the shocked material lies above, or at very shallow depths within, the
photosphere and this occurs only for accretion rates below 
$\sim$$10^{-8}$~\msun\ yr$^{-1}$.

The lower limit to the accretion rate  
 obtained by assuming all the accretion luminosity emerges as 
 X-radiation is only $\dot{M}_7$$\sim$6$\times$$10^{-5}$ or
 $\dot{M}$$\sim$6$\times$$10^{-12}$~\msun~yr$^{-1}$.
At such a low rate most of the accretion luminosity should be observed in 
 X-rays for plausible ranges of filling factor.
The actual accretion rate, however, must be higher
 as only a fraction of the available energy emerges in
 X-rays and a portion of the accretion energy undoubtably drives the 
 observed jets and \hh\ outflows from \hd.

This scenario also requires a stellar magnetic field.
For steady-state accretion,
 the disk is disrupted by the stellar magnetic field at a radius, $R_M$,
 roughly the corotation radius 
 where the Keplerian angular velocity in the disk equals the 
 angular velocity at the stellar surface (e.g., Shu \etal\ 1994).
For \hd, this corresponds to $R_M$$=$10.6$R_{\star}$ 
 (adopting $v$sin$i$$\sim$120~km~s$^{-1}$ as tabulated
 by van~den~Ancker \etal\ 1998).
This radius is where the torque exerted by the magnetic field on the disk
 approximately equals the viscous torque in the disk and is given by the
 Alfv\'{e}n radius, $r_A$, for a dipole field.
The Alfv\'{e}n radius can be expressed in terms of the magnetic field strength 
 at the stellar surface, the stellar mass, radius, and accretion rate
 (e.g., Frank, King, \& Raine 1985).
For \hd, $r_A$$=$7.2$\times$$10^9$$\dot{M}_7^{-2/7}$$B_{\star}^{4/7}$.
Assuming $r_A$$=$$R_M$ results in a value of 
 $B_{\star}$$\sim$$10$$\dot{M}_7^{1/2}$~kG for the surface field strength
 for \hd.

For the minimal accretion rate of
$\dot{M}_7$$\sim$6$\times$$10^{-5}$, the required magnetic field
strength is only $\sim$80~G.  
Higher magnetic fields, of order 500~G, have been invoked to explain
confinement of plasma during strong flares observed from the \hae\
star V892 (Giardino \etal\ 2004) and the Herbig~Be star MCW~297
(Hamaguchi \etal\ 2000).  Very recently, Hubrig, Sch\"{o}ller, \& Yudin
(2004) have presented evidence for (longitudinal) 
 fields of several hundred Gauss on Herbig~Ae
 stars~---~values consistent with surface field strengths of up to 
 2~kG measured from many classical T~Tauri stars (e.g., Guenther \etal\ 1999;
 Johns-Krull, Valenti, \& Koresko 1999).  
Theoretically, these are plausible relic
field strengths for a young star (see references in the Introduction) and
 imply accretion 
 rates of order $\dot{M}_7$$\sim$$10^{-3}$ or higher.
Combining this with the simplistic arguments based on 1-D shocks given above
 suggests an accretion
 rate of $\dot{M}$$\sim$$10^{-9\pm1}$~\msun~yr$^{-1}$, tuned
 such that the great majority of the X-ray emission is reprocessed within
 the photosphere and such that only of order 1~\%\ or less is actually
 observed at X-ray wavelengths.
This rate is comparable to those reported in the literature.
Gullbring \etal\ (1998) report rates as low as a few $10^{-10}$~\msun~yr$^{-1}$ for a number of T~Tauri stars while
Calvet \etal\ (2004) find rates of a few $10^{-8}$~\msun~yr$^{-1}$ for young 1--3~\msun\ T~Tauri (pre-\hae) stars.

The accretion rate and magnetic field strength of \hd\ can be better 
 determined when the contributions from the observed mass outflow are 
 properly accounted for in the overall energy budget. 
Until then we conclude that the X-ray emission from \hd\ is most
 consistent with an optically-thin shocked accretion flow confined by
 a stellar (dipole) magnetic field of unknown origin.
The X-ray data do not require a a late-type
companion be present, and are in fact inconsistent with the hotter
coronal temperatures ($10^7$~K) that typically characterize active
late-type stellar coronae.  A high accretion rate and strong magnetic
field are also not required, though would ease difficulties that are
otherwise present in explaining the accretion-driven X-rays using
simplistic accretion shock models.

\hd\ is remarkable in being in a class of a very few pre-Main Sequence
 stars whose
observed X-ray emission has such a cool characteristic temperature
that it can be entirely explained in terms of accretion. Accreting
T~Tauri stars generally exhibit X-ray emission characteristic of
plasma at tempertures of $10^7$~K or higher, similar to very active
late-type stars.  There is to our knowledge only one other PMS object
whose X-ray emission is also accretion-dominated: the nearby T~Tauri
star TW~Hya (e.g.\ Kastner et al.\ 2002). Other phenomena such as 
giant outbursts have also been observed (Kastner et al. 2004)
that signal the onset of mass accretion at rates much higher than 
required to explain the X-ray behavior reported here for HD 163296.

\subsection{X-ray Knot }

Detection of X-ray emission associated with \hh\ objects is not unprecedented.
These shock-heated
 objects are formed when supersonic outflow collides with ambient material or
 previous ejecta (e.g., Schwartz 1983).
Pravdo et al. (2001) report X-ray emission from HH2 in
 Orion at the level of $5\times 10^{29}$~\ergl\
 and Favata et al. (2002) estimate a
 luminosity of $\sim$$3\times 10^{29}$~\ergl\ from HH knots in the proto-stellar
 jet of L1551 IRS5.
The luminosity deduced here for the isolated knot in the SW jet
 is far less than these values yet is
 not unreasonable for an adiabatic shock. 
From the analytic model of Raga \etal\ (2002), the luminosity
 is
$L$$\sim$$1.8$$\times$$10^{29} ( n/100)^2 (r/10^{16})^3 (v/100)$~\ergl\
for a shock velocity $v$ km~s$^{-1}$, a pre-shock density $n$ cm$^{-3}$, and
a bow shock length scale $r$ cm.
For a shock
 velocity equal to that of the jet ($\sim$350 km/s), a typical pre-shock density
 ($\sim$100~cm$^{-3}$),
 and a bow shock length scale of order the optical size
 of the knot ($\sim$$10^{15}$ cm), the free-free luminosity is
 $\sim$$6\times 10^{26}$~\ergl. This is comparable to the value 
 of $\sim$3$\times$$10^{27}$~\ergl\ estimated
 here for the emission feature along the \hd\ jet axis.
 
\acknowledgements

Support for this research was provided in part by
NASA/\cxo\ Award Number GO3-4006X to DAS.
Optical data are
based on observations made with the Apache Point Observatory
3.5m telescope, which is owned and operated by the
Astrophysical Research Consortium. The Goddard Fabry-Perot (GFP)
is supported by NASA RTOP 188-01-22. Data analysis
facilities for the GFP are provided by the Laboratory for
Astronomy and Solar Physics at NASA's GSFC.
Use was also made of \hst\ observations obtained as part of HST-GTO-7065, 
HST-GTO-8065, and HST-GTO-8801. 
Support for the STIS IDT was provided by NASA Guaranteed Time Observer (GTO) 
funding to the STIS Science Team in response to NASA A/O OSSA -4-84 through 
the Hubble Space Telescope Project at Goddard Space Flight Center. 
CAG was supported through NASA PO 70789-G and NASA PR 4200048153 to Eureka 
Scientific.


\begin{thebibliography}{}
%
%
\bibitem[]{1606}
Appenzeller, I. 1994 in The Nature and Evolutionary Status of Herbig Ae/Be 
Stars, ASP Conf. Series 62, ed. P. S. Th\'{e}, M. R. Perez, E. P. J. van den Heuvel
%
\bibitem[]{1606}
Baade, D., \& Stahl, O. 1989, A\&A, 209, 268
%
\bibitem[]{1606}
Beskrovnaya, N. G., Pogodin, M. A., Yudin, R. V., Franco, G. A. P., Vieira, S. L. A., \& Evans, A. 1998, A\&AS, 127, 243
%
\bibitem[]{1606}
Birk, G. T., Schwab, D., Wiechen, H., \& Lesch, H. 2000, A\&A, 358, 1027
%
\bibitem[]{1607}
Bohlin, R. C., Savage, B. D., \& Drake, J. F. 1978, ApJ, 224, 132
%
\bibitem[]{1606}
Borkowski, K. J., Lyerly, W. J., \& Reynolds, S. P. 2001, ApJ, 548, 820
%
%
\bibitem[]{1606}
Brown, D. A., P\'{e}rez, M. R., \& Yusef-Zadeh, F. 1993, AJ, 106, 2000
%
\bibitem[]{1606}
Calvet, N., \& Gullbring, E. 1998, ApJ, 509, 802
%
\bibitem[]{1606}
Calvet, N., Muzerolle, J., Brice\~{n}o, C., Hern\'{a}ndez, J., Hartmann, L., Saucedo, J. L., \& Gordon, K. D. 2004, AJ, 128, 1294
%
\bibitem[]{1606}
Catala, C., Simon, T., Praderie, F., Talavera, A., Th\'{e}, P. S., \& Tijn A Djie, H. R. E. 1989, A\&A, 221, 273
%
\bibitem[]{1606}
Drake, J.J., 2004, in ``Cool Stars, Stellar Systems and the Sun: 13th
Cambridge Workshop'', eds.\ F.~Favata and G.~Hussain, in press
%
\bibitem[]{1606}
Devine, D., \etal\ 2000, ApJ, 542, L115
%
\bibitem[]{1606}
Dudorov, A. E., Krivodubskii, V. N., Ruzmaikina, T. V., \& Ruzmaikin, A. A. 1989, Sov. Astr., 33, 420
%
\bibitem[]{1606}
Favata, F., Fridlund, C. V. M., Micela, G., Sciortino, S., \& Kaas, A. 2002, A\&A, 386, 204
%
\bibitem[]{1606}
Feigelson, E. D., \& Montmerle, T. 1999, ARA\&A, 37, 363
%
\bibitem[]{1606}
Frank, J., King, A. R., \& Raine, D. J. 1985, Accretion power in Astrophysics (Cambridge: CUP)
%
\bibitem[]{1606}
Giardino, G., Favata, F., Micela, G., \& Reale, F. 2004, A\&A, 413, 669
%
\bibitem[]{1606}
Gilliland, R. L. 1986, ApJ, 300, 339
%
\bibitem[]{1606}
Grady, C. A., \etal\ 2000, ApJ, 544, 895
%
\bibitem[]{1606}
Grady, C. A., et al. 2004a, ApJ 608, 809 
%
\bibitem[]{1606}
Grady, C. A., Williger, G. M., Bouret, J. -C., Roberge, A., Sahu, M., 
Woodgate, B. 2004b, to appear in ``Astrophysics in
the Far Ultraviolet,'' ASP Conf. Ser., ed. B.-G. Anderssen, G. Sonneborn, W. Moos (in press)
%
\bibitem[]{1606}
Guenther, E. W., Lehmann, H, Emerson, J. P., \& Staude, J. 1999, A\&A, 341, 768
%
%
%
\bibitem[]{1606}
Hamaguchi, K., Koyama, K., Yamauchi, S., \& Terada, H. 2002, in Stellar Coronae in the Chandra and XMM-Newton Era, ASP Conf. Series 277, 193, ed. F. Favata and J. J. Drake
%
\bibitem[]{1606}
Hamaguchi, K., Terada, H., Bamba, A., \& Koyaman K. 2000, ApJ, 532, 1111
%
\bibitem[]{1606}
Hamaguchi, K., Yamauchi, S., \& Koyama, K. 2004, to be published in ApJ, astro-ph/0406489
%
%
\bibitem[]{1606}
Hayashi, M. R., Shibata, K., \& Matsumoto, R. 1996, ApJ, 468, L37
%
\bibitem[]{823}
Herbig, G. H. 1960, ApJS, 4, 337
%
\bibitem[]{1606}
Hillenbrand, L. A., Strom, S. E., Vrba, F. J., \& Keene, J. 1992, ApJ, 397, 613
%
\bibitem[]{1606}
Hubrig, S., Sch\"{o}ller, M., \& Yudin, R. V. 2004, astro-ph/0410571
%
\bibitem[]{1606}
Johns-Krull, C. M., Valenti, J. A., \& Koresko, C. 1999, ApJ, 516, 900
%
%
\bibitem[]{1606}
Kastner, J.~H., Huenemoerder, D.~P., Schulz, N.~S., Canizares, C.~R., \& Weintraub, D.~A.\ 2002, \apj, 567, 434
%
\bibitem[]{1606}
Kastner, J.~H., \etal\ 2004, Nature, 430, 429
%
\bibitem[]{1606}
K\"{o}nigl, A. 1991, ApJ, 370, L39
%
\bibitem[]{1606}
Lamzin, S. A. 1995, A\&A, 295, L20
%
\bibitem[]{1606}
Lamzin, S. A. 1999, Astron. Let. 25, 430
%
\bibitem[]{1606}
Linsky, J. L. 2003, AdSpR, 32, 917
%
\bibitem[]{1606}
Mannings, V., \& Sargent, A. I. 1997, ApJ, 490, 792
%
\bibitem[]{1606}
Muzerolle, J., D'Alessio, P., Calvet, N., \& Hartmann, L. 2004 astro-ph/0409008
%
%
\bibitem[]{1606}
Pagano, I., Linsky, J. L., Carkner, L., Robinson, R. D., Woodgate, B., 
 Timothy, G. 2000, ApJ 532, 497
%
\bibitem[]{1606}
Palla, F., \& Stahler, S. W. 1993, ApJ, 418, 414
%
\bibitem[]{1606}
Pogodin, M. A. 1994, A\&A, 282, 141
%
\bibitem[]{1606}
Popham, R., Narayan, R., Hartmann, L., \& Kenyon, S. 1993, ApJ, 415, L127
%
\bibitem[]{1606}
Pravdo, S. H., Feigelson, E. D., Garmire, G., Maeda, Y., Tsubol, Y., \& Bally, J. 2001, Nature, 413, 708
%
%
\bibitem[]{1606}
Raga, A. C., Noriega-Crespo, A., \& Velazquez, P. F. 2002, ApJ, 576, L149
%
\bibitem[]{1606}
Robinson, R.D., Linsky, J.L., Woodgate, B.E., Timothy, J.G. 2001, ApJ 554, 368
%
\bibitem[]{1606}
Romanova, M. M., Ustyugova, G. V., Koldoba, A. V., Chechetkin, V. M., \& Lovelace, R. V. E. 1998, ApJ, 500, 703
%
\bibitem[]{1606}
Schmitt, J. H. M. M. 1997, A\&A, 318, 215
%
\bibitem[]{1606}
Schwartz, R. D. 1983, ARA\&A, 21, 209
%
%
%
%
\bibitem[]{1606}
Shang, H, Glassgold, A. E., Shu, F. H., \& Lizano, S. 2002, ApJ, 564, 853
%
\bibitem[]{1606}
Shu, F., Najita, J., Ostriker, E., Wilkin, F., Ruden, S., \& Lizano, S. 1994, ApJ, 429, 781
%
\bibitem[]{1606}
Simon, T., Drake, S. A., \& Kim, P. D. 1995, PASP, 107, 1034
%
\bibitem[]{1606}
Stahler, S. W., Shu, F. H., \& Taam, R. E. 1980, ApJ, 241, 637
%
\bibitem[]{1606}
Th\'{e}, P. S., de~Winter, D., \& P\'{e}rez, M. R. 1985, A\&AS, 104, 315
%
\bibitem[]{1606}
Tout, C. A., \& Pringle, J. E. 1995, MNRAS, 272, 528
%
\bibitem[]{1606}
Ulrich, R. K. 1976, ApJ, 210, 377
%
\bibitem[]{1606}
van~den~Ancker, M. E., de~Winter, D., \& Tijn~A~Djie, H. R. E. 1998, A\&A, 330, 145
%
\bibitem[]{1606}
von~Rekowski, B., \& Brandenburg, A. 2004, A\&A 420, 17
%
\bibitem[]{1606}
Waters, L. B. F. M., \& Waelkens, C. 1998, ARA\&A, 36, 233
%
%
\end{thebibliography}
\end{document}